\documentclass[prb,aps,twocolumn,amsmath,amssymb,showpacs,superscriptaddress]{revtex4-1}
\usepackage{graphicx}
\DeclareGraphicsExtensions{.eps}
\usepackage[english]{babel}
\usepackage{natbib}
\usepackage{amsfonts}
\usepackage{amssymb}
\usepackage{dcolumn}
\usepackage{amsmath}
\usepackage[ansinew]{inputenc}
\usepackage{amssymb}
\usepackage{multirow}
\usepackage{mathtools}
\usepackage{wasysym}
\usepackage{color}
\usepackage{verbatim}
\usepackage{mciteplus}

\renewcommand{\vec}[1]{{\bf #1}}
\newcommand{\adim}[1]{\widetilde{#1}}

\newcommand{\sep}{\hspace{0.15cm}}

\begin{document}
\title{Critical behavior of plastic depinning of vortex lattices in two dimensions: Molecular dynamics simulations}

\author{Y. Fily}
\affiliation{LEMA, UMR 6157, Universit\'e F. Rabelais-CNRS-CEA, Parc de Grandmont, 37200 Tours, France}
\affiliation{Department of Physics, Syracuse University, Syracuse, NY 13244, USA}
\author{E. Olive}
\author{N. Di Scala}
\author{J.C. Soret}
\affiliation{LEMA, UMR 6157, Universit\'e F. Rabelais-CNRS-CEA, Parc de Grandmont, 37200 Tours, France}

\begin{abstract}
Using molecular dynamics simulations, we report a study of the dynamics of
two-dimensional vortex lattices driven over a disordered medium.
In strong disorder, when topological order is lost, we show that the depinning transition is analogous to a second order critical transition: the velocity-force response at the onset of motion is continuous and characterized by critical exponents. Combining studies at zero and nonzero temperature and using a scaling analysis, two critical exponents are evaluated. We find $v\sim (F-F_c)^\beta$ with $\beta=1.3\pm0.1$ at $T=0$ and $F>F_c$, and $v\sim T^{1/\delta}$ with $\delta^{-1}=0.75\pm0.1$ at $F=F_c$, where $F_c$ is the critical driving force at which the lattice goes from a pinned state to a sliding one.
Both critical exponents and the scaling function are found to exhibit universality with regard to the pinning strength and different disorder realizations.
Furthermore, the dynamics is shown to be chaotic in the whole critical region. 
\end{abstract}

\maketitle

\section{Introduction}

Due to the competition between interactions and randomness,
the dynamics of coherent structures driven over a disordered medium exhibit a great variety of phases.
In particular, the transition from a pinned state to a sliding one occurring at a critical driving force $F_c$, known as the \emph{depinning transition}, is 
both a great theoretical challenge and relevant for numerous systems:
superconductor vortices, colloids, Wigner crystals, magnetic bubbles, charge density waves, magnetic domain walls...
It has been suggested on phenomenological grounds by Fisher \cite{Fisher1985} that the depinning transition could be regarded as a critical phenomenon in which the velocity and the driving force would respectively be the order parameter and the control parameter.
Though it was originally intended to describe the depinning of charge density waves (CDW), this idea has proven very useful in many other domains.
In the elastic limit, it has been shown theoretically that for most manifolds and for CDW
the depinning does behave like a second order transition with a power law response $v\sim (F-F_c)^\beta$ at the onset of motion and $\beta<1$ \cite{Nattermann1992,*Narayan1992,*Narayan1993,*Ertas1994,*Chauve2000,*Chauve2001,*LeDoussal2002}.
When it comes to situations in which an elastic description is no longer valid, however, the theoretical description of the phenomenon is much more difficult.
In particular, the nature of the depinning transition (continuous or discontinuous) remains an open problem.
A continuous depinning transition (second order) with $\beta>1$ is observed in experiments \cite{Duruoz1995,*Rimberg1995,*Kurdak1998,*Parthasarathy2001,Higgins1996,*Ruyter2008,*Ammor2010,Mohan2009} and numerical simulations \cite{Dominguez1994a,*Chen2008a,*Chen2008b,*Liu2008,*Guo2009,*Lv2009,Reichhardt2001,Reichhardt2002,Reichhardt2003,Olive2009}. On the other hand, experiments on CDW \cite{Maeda1990}, and 3D numerical simulations of vortices \cite{Olson2001} suggest a discontinuous (first order) depinning transition where the velocity-force curve displays hysteresis and jumps between pinned and unpinned states.
On the theoretical side, various models have been developed to describe non elastic dynamics. 
A coarse grained model has been proposed \cite{Marchetti2000,*Marchetti2002,*Marchetti2002a,*Marchetti2003,*Marchetti2005} in which a visco-elastic coupling is used as an effective description of topological defects or phase slips. In the mean field limit, it predicts two kinds of depinning: a continuous one, belonging to the universality class of elastic depinning, and an hysteretic one.
The existence of an hysteretic depinning in a special case of this model has been confirmed using functional renormalization \cite{LeDoussal2008}.
Other phase-slip models also predict hysteresis for CDW \cite{Strogatz1988,*Levy1992}.
Conversely, numerical studies of a model focusing on the filamentary nature of the flow in 2D show a continuous depinning with $\beta\simeq1.5$ \cite{Watson1996,*Watson1997}.
Other phase-slip models based on the XY model also suggest the absence of hysteresis at the thermodynamic limit\cite{Nogawa2003} (in 3D) and a continuous second-order plastic depinning transition with an exponent $\beta\simeq1.7$\cite{Kawaguchi1999} (in 1D).
Another approach is to introduce inertia in the equation of motion \cite{Schwarz2001,*Schwarz2003,*Sengupta2010}. In this case a continuous depinning transition is found for small inertial parameter, otherwise a discontinuous transition with hysteresis is found.
In periodic systems with a displacement field of dimension $N=2$ (\emph{e.g.} superconductor vortices, colloids, Wigner crystals),
simulations have established that strong disorder leads to dislocations and plasticity. At the depinning threshold, regions of pinned particles (zero velocity) coexist with particles flowing around the pinned regions \cite{Jensen1988,*Jensen1988a,*Koshelev1992,*Ryu1996,*Moon1996,*Faleski1996,
*Spencer1997,*Olson1998,*Kolton1999,*Cao2000,*Fangohr2001,*Chen2003,*Chandran2003,
Groenbech-Jensen1996,Reichhardt2001,Reichhardt2002,
Olive2006,*Olive2008}.
Besides, the transition seems to be continuous and smooth ($\beta>1$) in 2D ($d=2$ is here the dimension of the embedding space) \cite{Reichhardt2001,Reichhardt2002,Reichhardt2003,Olive2009}.
Simulations on 2D colloids also indicate that the relaxation time near depinning obeys a power law, as expected in a second order transition \cite{Reichhardt2009}.
However,
most studies in which critical exponents are evaluated are carried out in the case of a 1D displacement field ($N=1$, \emph{e.g.} Josephson junction arrays, metallic dots, CDW), while there are very few such studies for $N=2$.
In particular, there is to the best of our knowledge no available study on superconductor vortices giving the $\beta$ exponent in the 2D plastic regime
(it should be noted though that $\beta$ has been numerically evaluated for superconductor vortices in 3D\cite{Luo2007}).

In this paper, we perform molecular dynamics simulations of 2D vortex lattices with strong random pinning and study the depinning transition induced by an external driving force.
Our system belongs to the category of 2D periodic systems with a 2D displacement field and short range interactions.
It could model 3D superconductors (either conventional or layered) in an effective 2D regime, \emph{i.e.} when the vortex line tension is high enough for the lines to remain straight.

The behavior of the system near the depinning transition is studied at both zero and nonzero temperature, which allows an accurate measurement of the critical driving force.
At zero temperature, the depinning is continuous and highly plastic.
The static channels observed at the onset of motion are identified with the so-called \emph{single particle regime} and considered as a finite size effect.
In the intermediate range of driving force, the motion is chaotic, whereas at high driving force the lattice reorders and chaos disappears.
In the threshold vicinity, a study of the temperature dependence of the velocity allows us to determine the true critical force despite the single particle regime.
Moreover, it is shown that the velocity scales as a power law of temperature $v_{F=F_c}\sim T^{1/\delta}$ at $F=F_c$ and the associated critical exponent $\delta$ is determined.
At $T=0$, a second power law $v\sim(F-F_c)^\beta$ is found with an exponent $\beta>1$.
Varying simultaneously the driving force and the temperature, we then find evidence of the existence of a scaling law, which confirms the values of $\beta$ and $\delta$ evaluated respectively at $T=0$ and $F=F_c$.
This analysis has been performed for various system sizes, disorder realizations and pinning strengths, indicating that both $\beta$, $\delta$ and the scaling function exhibit some degree of universality.

\section{Numerical model}

\bibpunct{}{}{,}{n}{}{}
As in Ref.\ \cite{Olive2009}, we study $N_v$ Abrikosov vortices interacting with $N_p$ random pins in the $(x,y)$ plane. We consider the London limit $\lambda_L\gg\xi$, where $\lambda_L$ is the penetration length and $\xi$ is the coherence length, \emph{i.e.} we treat vortices as point particles. The overdamped equation of motion of a vortex $i$ at position ${\bf r}_i$ reads
\bibpunct{}{}{,}{s}{}{}
\begin{equation}
\begin{split}
\eta \frac{d{\bf r}_i}{dt}=- & {\sum_{j \neq i}}\nabla_i U^{vv}(r_{ij})-{\sum_{p}}\nabla_i U^{vp}(r_{ip})  \\
  & +{\bf F}^L+{\bf F}_i^{\text{th}}(t)
\end{split}
\end{equation}
\noindent
where $r_{ip}$ is the distance between the vortex $i$ and the pinning site located at ${\bf r}_p$, $r_{ij}$ is the distance between the vortices $i$ and $j$ located at ${\bf r}_i$ and ${\bf r}_j$,
and $\nabla_i$ is the 2D gradient operator acting on ${\bf r}_i$.
$\eta$ is the viscosity coefficient. ${\bf F}^L=F{\bf \hat x}$ is the Lorentz driving force due to an applied current. ${\bf F}_i^{\text{th}}$ is the thermal gaussian white noise with zero mean and variance
$$\langle F_{i,\mu}^{\text{th}}(t) F_{j,\nu}^{\text{th}}(t') \rangle=2 \eta k_B T \delta_{ij} \delta_{\mu \nu} \delta (t-t')$$
where $\mu,\nu=x,y$ and $k_B$ is the Boltzmann constant.
The vortex-vortex repulsive interaction is given by a modified Bessel function
$$U^{vv}(r_{ij})=\alpha_v K_0(r_{ij}/\lambda_L)$$
and the attractive pinning potential is given by
$$U^{vp}( r_{ip})=-{\alpha_p} e^{-(r_{ip}/R_p)^2}$$
where $R_p$ the radius of the pins, and $\alpha_v$ and $\alpha_p$ are tunable parameters.
Depending on the relative strengths of the vortex-vortex and vortex-pin interactions, the dynamics can be either dominated by elasticity or disorder.
The strength of the vortex-vortex interaction is fixed by setting $\alpha_v=2.83\ 10^{-3}\lambda_L \epsilon_0$ where $\epsilon_0$ is an energy per unit length. The relative disorder strength $\alpha_p/\alpha_v$ is then chosen high enough for the depinning transition to exhibit plasticity (in our model, plasticity is found above $\alpha_p/\alpha_v\sim0.01$).
Molecular dynamics simulation is used for $N_v=30 n^2$ vortices in a rectangular basic cell $(L_x,L_y)=(5, 6 \sqrt3/2) n \lambda_L$, with $n$ from $3$ to $8$, \textit{i.e.} $N_v$ from $270$ to $1080$.
Periodic boundary conditions are used in both directions. The vortex-vortex interaction is dealt with using a neighbor list method with a cutoff radius $r_c=7.1\ \lambda_L$.
The number of pins is set to $N_p=5N_v$, and their radius is $R_p=0.22\ \lambda_L$. The average vortex distance is 
$a_0=\lambda_L $.
We use a unit system in which $\eta=1$, $\lambda_L=1$, $\epsilon_0=1$ and $k_B=1$. \\

\section{Behavior at $T=0$}
\label{T=0}
\begin{figure}[!h]
\includegraphics[width=0.6\linewidth]{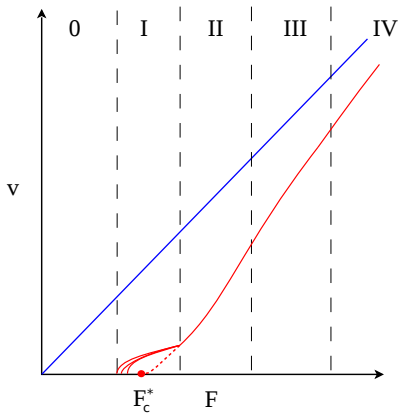}
\caption{(color online) Schematic of the average velocity $v$ versus driving force $F$. The vertical dashed lines separate the different kinds of flow observed: pinned static lattice (region 0), single particle regime (region I), disordered chaos (region II), smectic chaos (region III) and decoupled channels (region IV).
The dotted line crossing the horizontal axis at $F=F_c^*$ is the prolongation of the curve from region II to region I (see text section \ref{delta}).
}
\label{schemaV-F}
\end{figure}

\begin{figure}[!h]
\includegraphics[width=\linewidth]{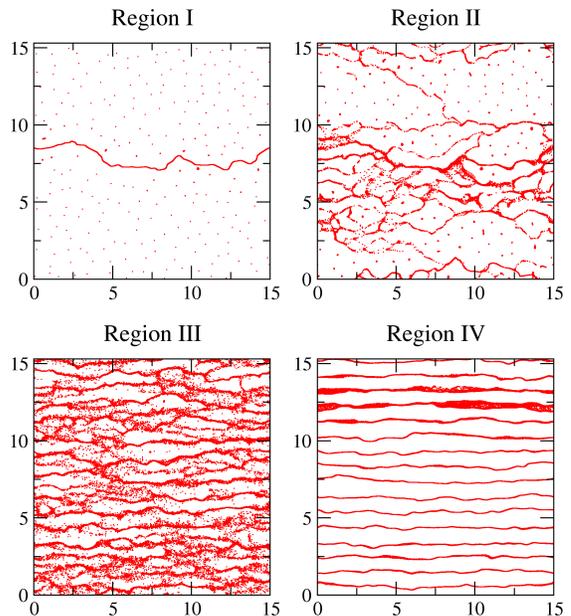}
\caption{(color online) Typical trajectories of the vortices in the four moving regions shown in Fig.~\ref{schemaV-F}: single particle regime (region I), disordered chaos (region II), smectic chaos (region III) and decoupled channels (region IV).}
\label{trajec}
\end{figure}

We start with a perfect lattice at high velocity and slowly decrease the driving force until the system reaches a pinned state. The force is then varied back and forth with various force steps in order to check for hysteresis. The whole process is done at $T=0$. In Fig.~\ref{schemaV-F}
we plot the typical shape of the average velocity-force curve, showing four distinct regions.
These regions are better illustrated by the typical trajectories of vortices\cite{Groenbech-Jensen1996,Reichhardt2001} displayed in Fig.~\ref{trajec}.

Just above the depinning threshold (region I), the flow occurs along one or several non crossing static channels while the rest of the system remains pinned (see Fig.~\ref{trajec}).
The velocity-force curve exhibits jumps and hysteresis.
This is related to the existence of several sets of such channels for a given driving force. Each set corresponds to an hysteresis branch while jumps in velocity are in fact jumps from a branch to another one, \emph{i.e.} from a set of channels to another one.
Note that the branch chosen by the system depends on the force ramping rate.
Moreover, on a given channel, the positions of vortices are fully determined by the position of one of them.
Indeed, it can be verified by choosing one vortex at a given position on the channel, and plotting the positions of all the other vortices.
Since the channel is one dimensional, this means that the dynamics can be modeled by a single degree of freedom.
Because of the periodic boundary conditions, this single degree of freedom sees a periodic potential, leading to the so-called \emph{single particle regime}: the system experiences a saddle-node bifurcation at the critical force ; above $F_c$, the velocity is a periodic function of time and scales as $v\sim (F-F_c)^{1/2}$.
We measure indeed this power law on each branch corresponding to a given configuration of the channels.
Moreover, simulations of boxes shaped as long strips in the longitudinal direction show that the range of force in which region I is observed decreases when the longitudinal size is increased. At large sizes, the size of region I seems to shrink to zero.
As a result, we assume that region I vanishes in the infinite size limit, and should not be taken into consideration for the study of the critical behavior.

In region II and III, vortices flow along changing interconnected channels  (see Fig.~\ref{trajec}). The transition from region II to region III is defined by the appearance of transverse smectic order indicated by small peaks in the structure factor along the $k_y$ axis, while the sketch of the high velocity channels becomes visible. It occurs near the inflexion point in the velocity-force curve, also known as the \emph{peak in differential resistance} in reference to the experimental tension-intensity curves.
In region IV, vortices stop switching channels (see Fig.~\ref{trajec}) and a linear behavior $v\sim F-F_c$ is observed. \\

\begin{figure}[h]
\includegraphics[width=\linewidth]{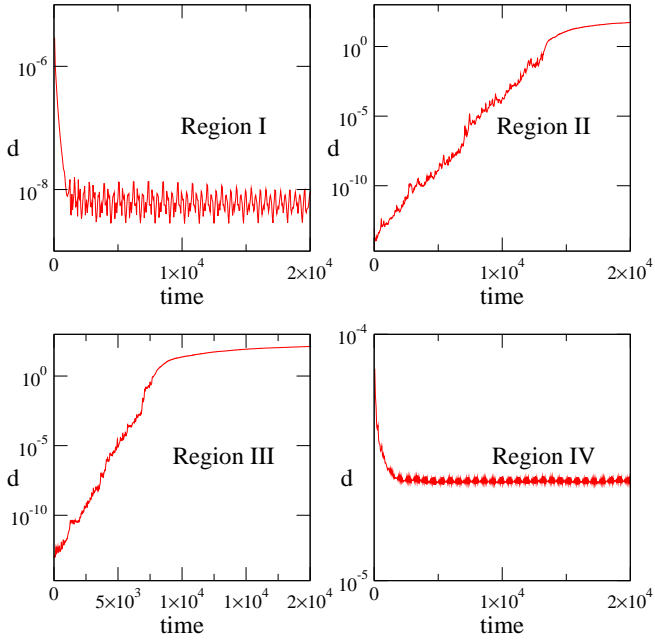}
\caption{(color online) Distance $d$ between two neighboring trajectories versus time in the region I, II, III and IV.}
\label{figlyap}
\end{figure}

\bibpunct{}{}{,}{n}{}{}
Following Ref.\ \cite{Olive2006,*Olive2008}, we now show that the motion in regions II and III is chaotic by evaluating the sign of the maximum Lyapunov exponent $\lambda$.\bibpunct{}{}{,}{s}{}{}
The existence of at least one positive Lyapounov exponent illustrates the sensitive dependence on initial conditions, which is a signature of chaos.
In Fig.~\ref{figlyap}, we plot the distance
\begin{align}
d(t)=\sqrt{\sum_{i=1}^{N_v}\left|{\bf r}_i^1(t)-{\bf r}_i^2(t)\right|^2}
\end{align}
in the phase space between two neighboring trajectories $\left(\vec{r}_{\scriptscriptstyle 1}^1(t),...,\vec{r}_{\scriptscriptstyle N_v}^1(t)\right)$ and $\left(\vec{r}_{\scriptscriptstyle 1}^2(t),...,\vec{r}_{\scriptscriptstyle N_v}^2(t)\right)$.
Two distinct behaviors are found.
In regions II and III, $d(t)$ grows exponentially at first and then saturates. The exponential growth indicates the existence of at least one positive Lyapunov exponent, which proves the existence of chaos. A saturation effect appears when $d(t)$ becomes of the order of the size of the chaotic attractor.
On the other hand, in regions I and IV $d(t)$ 
remains constant at large times (the largest Lyapunov exponent is zero) indicating a closed orbit in phase space while the decrease at short times is due to the transient regime associated with the existence of negative Lyapunov exponents.

\section{Critical exponents and scaling}
\label{critic}
In the original approach of Fisher \cite{Fisher1985}, it is argued that the depinning transition is a critical phenomenon implying scaling laws and critical exponents near threshold. It is therefore expected that $v_{F=F_c}\sim T^{1/\delta}$, and $v_{T=0,F>F_c}\sim (F-F_c)^\beta$, where $\delta$ and $\beta$ are critical exponents. In the following, we show that our data support the existence of such a critical phenomenon: the critical exponents $\delta$ and $\beta$ are determined, and a scaling relation is found between the velocity, temperature and driving force.
Note that the threshold behavior that we observe in our simulations is effectively continuous since hysteresis and jumps have only been observed in region I, which we believe vanishes in the infinite size limit.

	\subsection{Critical force and exponent $\delta$}
	\label{delta}

		\begin{figure}[h]
		\includegraphics[width=0.85\linewidth]{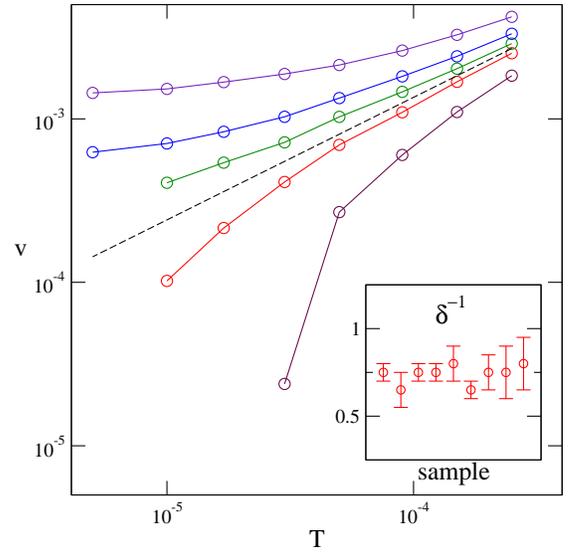}
		\caption{(color online) Average velocity $v$ versus temperature $T$ at low temperature for different forces around the critical force $F_c^*$ (from bottom to top: $10^3\times F=11 ; 12 ; 12.5 ; 13 ; 14$), $\alpha_p/\alpha_v=1.05$ and $N_v=270$. The dashed black line is the extrapolation of $v(T)$ at the critical force from which we extract $\delta^{-1}$.
		Inset: value of $\delta^{-1}$ for different samples:
$2$ different disorder strengths $\alpha_p/\alpha_v=1.05$ and $\alpha_p/\alpha_v=0.35$, $3$ different sizes ($N_v=270$ for $\alpha_p/\alpha_v=1.05$, $N_v=270$, $480$ and $1080$ for $\alpha_p/\alpha_v=0.35$), and different sets of positions of the pins ($2$ sets for $\alpha_p/\alpha_v=1.05$ and $N_v=270$, $5$ sets for $\alpha_p/\alpha_v=0.35$ and $N_v=270$).
		The error bars correspond to the different lines one can draw to extrapolate a power law behavior at the change of convexity.
		}
		\label{figscalingT}
		\end{figure}

First of all, we want to determine the critical force $F_c$.
The main issue is that because of hysteresis in region I, there is no unique depinning force $F_c$ directly accessible to measurements. Consequently, we have to evaluate an effective critical force $F_c^*$ by prolongating the velocity-force curve from region II to region I. The intersection
with the $v=0$ axis defines $F_c^*$ (see Fig.~\ref{schemaV-F}).
This evaluation can be improved by studying the temperature dependence of the velocity close to the depinning threshold.
As shown in Fig.~\ref{figscalingT}, two different behaviors are observed, depending on which side of the transition the force is.
Above the critical force, $v$ approaches a non zero limit as $T$ goes to $0$, leading to convex curves with an horizontal asymptote on the left in logarithmic scale.
Below the critical force, $v$ goes to $0$ faster than a power law, resulting in concave curves in logarithmic scale.
The change in convexity when the force is varied indicates that the effective critical force $F_c^*$ has been crossed.
The results obtained by these two methods are consistent, and combining both allows to improve the accuracy and precision of $F_c^*$.

	Furthermore, in agreement with a second order phase transition, we can extrapolate at $F=F_c^*$ a power law response \cite{Fisher1985} (a linear response in logarithmic scale at the change of convexity) from which we measure the critical exponent $\delta$:
\begin{equation}
v_{F=F_c^*}\sim T^{1/\delta}
\end{equation}
	As we shall see in section \ref{scaling}, this is consistent with the existence of a scaling relation, and the value of $\delta$ obtained here allows to collapse all data available on a single curve, supporting the validity of extrapolating a power law behavior at  $F=F_c^*$.
		This analysis has been performed on $9$ samples with different pinning strengths ($\alpha_p/\alpha_v=1.05$ and $0.35$), different system sizes ($N_v=270$ to $1080$) and different realizations of the random positions of the pinning centers.
		 The resulting values of $\delta^{-1}$ are shown in the inset of Fig.~\ref{figscalingT}, leading to $\delta^{-1}=0.75\pm 0.1$.
		
	\subsection{Critical exponent $\beta$}
	\label{beta}

\begin{figure}[h!]
\includegraphics[width=0.85\linewidth]{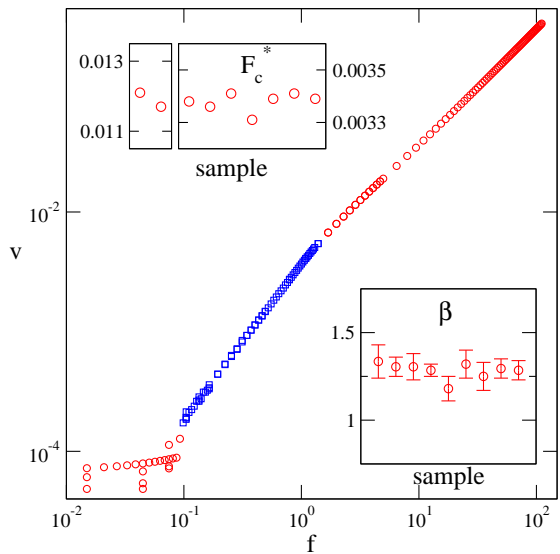}
\caption{(color online) Average velocity $v$ versus effective reduced driving force $f=(F-F_c^*)/F_c^*$ (circles and squares), including the critical region (squares).
Bottom right inset: value of $\beta$ for different samples (same samples as Fig.~\ref{figscalingT}: $\alpha_p/\alpha_v=1.05$ and $0.35$, $N_v=270$ to $1080$ and different sets of positions of the pinning centers).
Top left inset: values of $F_c^*$ for different samples (left side: $\alpha_p/\alpha_v=1.05$, $2$ samples ; right side: $\alpha_p/\alpha_v=0.35$, $7$ samples).}
\label{figscalingf}
\end{figure}

	We now go back to the $T=0$ case to study the velocity-force curve using the values of $F_c^*$ obtained in the previous section.	
	The mean velocity $v$ versus effective reduced force $f=(F-F_c^*)/F_c^*$ is plotted in Fig.~\ref{figscalingf} for $\alpha_p/\alpha_v=0.35$, showing the existence of a power law regime in the critical region, which lies from the lower boundary of region II ($f\sim0.1$) to close to its upper boundary ($f\sim 1$)
\begin{equation}
v_{T=0,f>0}\sim f^\beta
\end{equation}
	This power law results in a linear region in logarithmic scale, which slope $\beta$ has been measured on the $9$ samples of section \ref{delta}.
No significant differences between the samples were measured, leading to a unique value $\beta=1.3\pm0.1$ (see bottom right inset of Fig. \ref{figscalingf}). The precision of $\beta$ is limited by the precision of $F_c^*$.
	Note that using only the $T=0$ results (see section~\ref{delta}) to determine the critical force gives a similar result, except that the uncertainty on $\beta$ is greater.
		
	\subsection{Scaling law}
		\label{scaling}
		The power law dependence of $v$ versus both $f$ and $T$ strongly suggests to go on with the identification of the depinning transition with a critical phenomenon and to look for evidence of a scaling relation between the velocity, driving force and temperature.
		First of all, we want this relation to be expressed in terms of dimensionless quantities, and to be independent of the prefactors in the two power laws mentioned before. We thus define $v_0$ and $T_0$ such as
		\begin{align}
		v_{f>0,T=0}=v_0 f^\beta \quad,\quad v_{f=0}= v_0 \left(\frac{T}{T_0}\right)^{1/\delta}
		\end{align}
		Considering dimensionless velocity $\adim{v}=v/v_0$ and temperature $\adim{T}=T/T_0$, we define the scaling function $S$ as follows:
		\begin{align}
		\adim{v} |f|^{-\beta} =S_\pm\left(\adim{T} |f|^{-\beta\delta}\right)
		\end{align}
		where the unknown branches $S_+$ and $S_-$ of the scaling function correspond respectively to $f>0$ and $f<0$.
		Moreover, the observed power law dependences $\adim{v}_{f>0,T=0}=f^\beta$ and $\adim{v}_{f=0}={\adim{T}}^{1/\delta}$ imply that $S_\pm(x)$ satisfies
		\begin{align}
		\lim_{x\rightarrow0} S_+(x)=1 \quad,\quad
		\lim_{x\rightarrow\infty}x^{-1/\delta} S_\pm(x)=1
		\label{ansatz}
		\end{align}
		Graphically, it means (in logarithmic scale) that $S_+$ is asymptotic to the horizontal axis for $T\rightarrow0$ (driving dominated regime) while both $S_+$ and $S_-$ have an oblique asymptote with slope $\delta^{-1}$ for $f\rightarrow0$ (temperature dominated regime).
		The intersection of these two asymptotes occurs at $x=1$, defining a force dependent crossover temperature $T=T_0 |f|^{\beta\delta}$ between the two regimes.
		Note that the change of variables $(v,T)\rightarrow(\adim{v},\adim{T})$ is equivalent to choosing the intersection of the asymptotes as the origin of coordinates.

\begin{figure}[!h]
\includegraphics[width=0.85\linewidth]{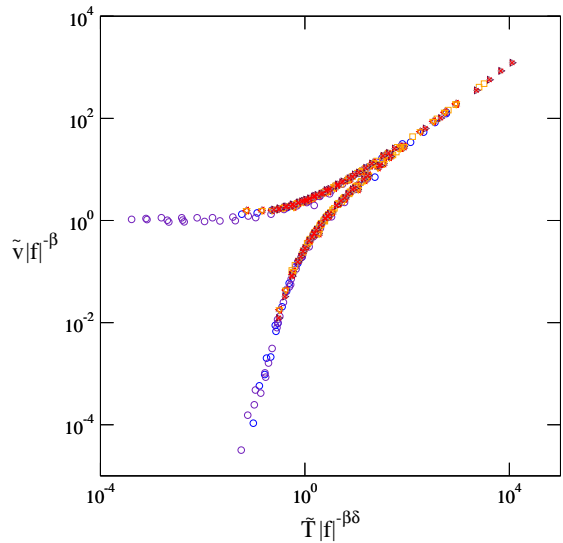} \\
\caption{(color online) Scaling plot $\adim{v} |f|^{-\beta}$ versus $\adim{T} |f|^{-\beta\delta}$ with $\beta=1.3$ and $\delta^{-1}=0.75$ featuring the $9$ samples shown in the insets of Fig.~\ref{figscalingT} and Fig.~\ref{figscalingf} ($\alpha_p/\alpha_v=1.05$ and $0.35$, $N_v=270$ to $1080$ and different sets of positions of the pinning centers).
}
\label{figscaling}
\end{figure}

We check for the existence of the scaling by plotting in Fig.~\ref{figscaling} $\adim{v} |f|^{-\beta}$ versus $\adim{T} |f|^{-\beta\delta}$. We observe a collapsing of data to a single curve (with two branches) for the same values of $F_c^*$, $\beta$ and $\delta$ obtained in sections \ref{delta} and \ref{beta}.
To be more specific, the data from all the samples are plotted using the same couple of values for $\beta$ and $\delta$, chosen equal to the average values obtained in previous sections ; $F_c^*$ on the other hand varies from sample to sample as shown in the inset of Fig.~\ref{figscalingf}.
For a given disorder strength $v_0$ and $T_0$ are constant, whatever the system size ($N_v=270$ to $1080$) and the positions of the pins.
When the disorder strength is changed, $v_0$ and $T_0$ change ($v_0=3.3\ 10^{-3}$ and $T_0=8\ 10^{-4}$ for $\alpha_p/\alpha_v=0.35$ ; $v_0=1.3\ 10^{-2}$ and $T_0=1.4\ 10^{-3}$ for $\alpha_p/\alpha_v=1.05$) but $S_\pm$ remains unchanged.

The collapsing of all the available data indicates that:
(i) the size effects are not relevant, \emph{i.e.} the system studied is large enough for a true critical regime to be observed
(ii) neither the critical exponents $\beta$ and $\delta$ nor the scaling function $\adim{S}_\pm$ depend on the disorder (strength and positions of the pins).
Our results therefore show some degree of universality within our model. The question or larger universality classes for plastic depinning of periodic objects with $N=2$ displacement fields will be addressed in the discussion.

\section{Discussion}

A large variety for the values of the $\beta$ exponent can be found in the literature, and the question of the existence of universality classes is still open for the plastic depinning transition. In some cases no scaling fit can even been found or hysteresis suggesting first order depinning transition has been reported. Our present results strongly suggest a second order depinning transition with well identified critical exponents $\beta$ and $\delta$ and scaling relations exhibiting some universality with regard to the disorder strengths, and disorder realizations are found. We therefore try to compare our results and in particular the depinning critical exponents to those reported in similar studies.

In the case of a displacement field of dimension $N~=~1$ (\emph{e.g.} Josephson junction arrays, XY model, metallic dots, CDW) in an embedding space of dimension $d=2$, the value $\beta=5/3$ has been predicted theoretically \cite{Middleton1993b}. Moreover, many studies are available, both experimental and numerical, and a large variety of $\beta$ values can be found (\emph{e.g.} $1.4<\beta <2.25$ for arrays of metallic dots \cite{Duruoz1995,*Rimberg1995,*Kurdak1998,*Parthasarathy2001},
and $1.3<\beta <2.6$ for Josephson junction and XY model
\cite{Dominguez1994a,*Chen2008a,*Chen2008b,*Liu2008,*Guo2009,*Lv2009}).

In the case of a displacement field of dimension $N~=~2$ (\emph{e.g.} superconductor vortices, colloids, Wigner crystals), there are to the best of our knowlegde only a few numerical studies proposing critical exponents for the plastic depinning transition (see Table \ref{tablexp}).%
\bibpunct{}{}{,}{n}{}{}%
Ref.~\cite{Reichhardt2003} uses a $N=2$ model to describe arrays of metallic dots, and the nature of the disorder's randomness is different from the other studies (regular array of pins with random strength vs. identical pins with random positions).%
\bibpunct{}{}{,}{s}{}{}%
The same study also considers the case of random positions by removing random sites on the array, however no unique value of $\beta$ is measurable in that case.
In Wigner crystals and colloids, $\beta$ is found to be
independent of the pinning strength \cite{Reichhardt2001,Reichhardt2002}.
In 3D superconductors ($d=3$) with well defined vortex lines, it is shown that one can measure two critical exponents $\beta$ and $\delta$, both independent of the disorder strength, and that a scaling relation involving these two exponents exists \cite{Luo2007}.
However, an hysteretic discontinuous transition is found if the superconductor is layered and the layers are allowed to decouple, \emph{i.e.} when there are no more vortex lines \cite{Olson2001}.
Finally in experiments on depinning in superconductors, $\beta \sim 1.2$ or $1.4$ is found above the peak effect \cite{Higgins1996,*Ruyter2008,*Ammor2010}, where the depinning is considered to be plastic.\\

\begin{widetext}
\begin{center}
\bibpunct{}{}{,}{n}{}{}
\renewcommand{\arraystretch}{1.3}
\begin{table}[h!]
\center
\begin{tabular}{| @{\sep} c @{\sep} | @{\sep} c @{\sep} | @{\sep} c @{\sep} | @{\sep} c @{\sep} | @{\sep} c @{\sep} | @{\sep} c @{\sep} |}
\hline
	\multirow{2}{*}{Study}		&	\multirow{2}{*}{\quad d \quad}	&
	particle-particle			&	\multirow{2}{*}{pinning}		&
	\multirow{2}{*}{$\beta$}	&	\multirow{2}{*}{$\delta^{-1}$}	\\
								&									&
	interaction					&									&
								&									\\ \hline \hline

\multirow{2}{*}{Ref.~\cite{Reichhardt2001}}	& 	\multirow{2}{*}{2}	&
\multirow{2}{*}{Coulomb}		&	Coulomb							&
$1.61\pm0.10$					&									\\
								& 									&
								&	random positions				&
$1.71\pm0.10$					&									\\ \hline\multirow{2}{*}{Ref.~\cite{Reichhardt2002}}	&  	\multirow{2}{*}{2}	&
\multirow{2}{*}{screened Coulomb}	&	parabolic					&
\multirow{2}{*}{$1.94\pm0.03$}	&									\\
								& 									&
								&	random positions				&
								&									\\ \hline
\multirow{2}{*}{Ref.~\cite{Reichhardt2003}}	& 	\multirow{2}{*}{2}	&
\multirow{2}{*}{Coulomb}		&	parabolic						&
\multirow{2}{*}{$1.94\pm0.15$}	&									\\
								& 									&
								&	random strength					&
								&									\\ \hline
\multirow{2}{*}{Ref.~\cite{Luo2007}}& 	\multirow{2}{*}{3}			&
\multirow{2}{*}{modified Bessel}&	gaussian						&
\multirow{2}{*}{$0.754\pm0.010$}& \multirow{2}{*}{$0.754\pm0.010$}	\\
								& 									&
								&	random positions				&
								&									\\ \hline
\multirow{2}{*}{Our results}	&  	\multirow{2}{*}{2}				&
\multirow{2}{*}{modified Bessel}&	gaussian						&
\multirow{2}{*}{$1.3\pm0.1$}	&	\multirow{2}{*}{$0.75\pm0.1$}	\\
								& 									&
								&	random positions				&
								&									\\ \hline
\end{tabular}
\caption{Depinning exponents (simulation results) in the plastic case for periodic systems with a displacement field of dimension $N=2$. $d$ is the dimension of the embedding space.}
\label{tablexp}
\end{table}
\bibpunct{}{}{,}{s}{}{}
\end{center}
\end{widetext}

In the present paper we propose (to the best of our knowledge) the first determination of both exponents $\beta$ and $\delta$, along with evidence of a scaling relation, for plastic depinning in the case $N=2$ and $d=2$ by studying effective 2D vortex lines.
First, we note that the value $\beta=1.3\pm 0.1$ is in agreement with vortex experiments above the peak effect \cite{Higgins1996,*Ruyter2008,*Ammor2010}. Second, 
we find some degree of universality with regard to the pinning strength and different realizations of the disorder, in agreement with previous results \cite{Reichhardt2001,Reichhardt2003}.
However, the variety of the values of the $\beta$ exponent reported above suggests that the type of particle-particle interaction as well as the type of disorder plays an important role. In the light of these results, it seems therefore difficult to define large universality classes for the plastic depinning of periodic objects in 2D.

\section{Conclusion}

In this paper, we studied the plastic depinning of vortex lattices in 2D
with  strong disorder. Above the pinned region, we  find four dynamical
regimes: periodic, disordered chaotic, smectic chaotic and decoupled
channels. The first one is controlled by the finite size of the simulation
box,  leading to the so-called \emph{single particle regime}.
A continuous (second order) phase transition is found at the depinning
threshold. The critical regime associated to the depinning transition is
found to be chaotic.
A careful analysis at $T=0$ and finite temperature allows an accurate determination of the critical force.
Scaling laws for the force and temperature
dependence of the velocity are found and two critical exponents are 
determined. Both critical exponents and the scaling function are independent 
of the disorder (strength and positions of the pins) in the range of 
parameters we have studied, indicating some degree of universality. However, the comparision with similar systems may suggest that large universality classes do not exist for the plastic depinning in 2D.

\section*{Acknowledgement}

We wish to thank Pierre Le Doussal, Alberto Rosso, Cristina Marchetti and Alan Middleton for helpful discussions.

\bibliographystyle{apsrevM}
\bibliography{biblio}

\ifx\mcitethebibliography\mciteundefinedmacro
\PackageError{apsrevM.bst}{mciteplus.sty has not been loaded}
{This bibstyle requires the use of the mciteplus package.}\fi
\begin{mcitethebibliography}{62}
\expandafter\ifx\csname natexlab\endcsname\relax\def\natexlab#1{#1}\fi
\expandafter\ifx\csname bibnamefont\endcsname\relax
  \def\bibnamefont#1{#1}\fi
\expandafter\ifx\csname bibfnamefont\endcsname\relax
  \def\bibfnamefont#1{#1}\fi
\expandafter\ifx\csname citenamefont\endcsname\relax
  \def\citenamefont#1{#1}\fi
\expandafter\ifx\csname url\endcsname\relax
  \def\url#1{\texttt{#1}}\fi
\expandafter\ifx\csname urlprefix\endcsname\relax\def\urlprefix{URL }\fi
\providecommand{\bibinfo}[2]{#2}
\providecommand{\eprint}[2][]{\url{#2}}

\bibitem[{\citenamefont{Fisher}(1985)}]{Fisher1985}
\bibinfo{author}{\bibfnamefont{D.~S.} \bibnamefont{Fisher}},
  \bibinfo{journal}{Phys. Rev. B} \textbf{\bibinfo{volume}{31}},
  \bibinfo{pages}{1396} (\bibinfo{year}{1985})\relax
\mciteBstWouldAddEndPuncttrue
\mciteSetBstMidEndSepPunct{\mcitedefaultmidpunct}
{\mcitedefaultendpunct}{\mcitedefaultseppunct}\relax
\EndOfBibitem
\bibitem[{\citenamefont{Nattermann et~al.}(1992)\citenamefont{Nattermann,
  Stepanow, Tang, and Leschhorn}}]{Nattermann1992}
\bibinfo{author}{\bibfnamefont{T.}~\bibnamefont{Nattermann}},
  \bibinfo{author}{\bibfnamefont{S.}~\bibnamefont{Stepanow}},
  \bibinfo{author}{\bibfnamefont{L.-H.} \bibnamefont{Tang}}, \bibnamefont{and}
  \bibinfo{author}{\bibfnamefont{H.}~\bibnamefont{Leschhorn}},
  \bibinfo{journal}{Journal de Physique II} \textbf{\bibinfo{volume}{2}},
  \bibinfo{pages}{1483} (\bibinfo{year}{1992})\relax
\mciteBstWouldAddEndPuncttrue
\mciteSetBstMidEndSepPunct{\mcitedefaultmidpunct}
{\mcitedefaultendpunct}{\mcitedefaultseppunct}\relax
\EndOfBibitem
\bibitem[{\citenamefont{Narayan and Fisher}(1992)}]{Narayan1992}
\bibinfo{author}{\bibfnamefont{O.}~\bibnamefont{Narayan}} \bibnamefont{and}
  \bibinfo{author}{\bibfnamefont{D.~S.} \bibnamefont{Fisher}},
  \bibinfo{journal}{Phys. Rev. B} \textbf{\bibinfo{volume}{46}},
  \bibinfo{pages}{11520} (\bibinfo{year}{1992})\relax
\mciteBstWouldAddEndPuncttrue
\mciteSetBstMidEndSepPunct{\mcitedefaultmidpunct}
{\mcitedefaultendpunct}{\mcitedefaultseppunct}\relax
\EndOfBibitem
\bibitem[{\citenamefont{Narayan and Fisher}(1993)}]{Narayan1993}
\bibinfo{author}{\bibfnamefont{O.}~\bibnamefont{Narayan}} \bibnamefont{and}
  \bibinfo{author}{\bibfnamefont{D.~S.} \bibnamefont{Fisher}},
  \bibinfo{journal}{Phys. Rev. B} \textbf{\bibinfo{volume}{48}},
  \bibinfo{pages}{7030} (\bibinfo{year}{1993})\relax
\mciteBstWouldAddEndPuncttrue
\mciteSetBstMidEndSepPunct{\mcitedefaultmidpunct}
{\mcitedefaultendpunct}{\mcitedefaultseppunct}\relax
\EndOfBibitem
\bibitem[{\citenamefont{Erta\c{s} and Kardar}(1994)}]{Ertas1994}
\bibinfo{author}{\bibfnamefont{D.}~\bibnamefont{Erta\c{s}}} \bibnamefont{and}
  \bibinfo{author}{\bibfnamefont{M.}~\bibnamefont{Kardar}},
  \bibinfo{journal}{Phys. Rev. E} \textbf{\bibinfo{volume}{49}},
  \bibinfo{pages}{R2532} (\bibinfo{year}{1994})\relax
\mciteBstWouldAddEndPuncttrue
\mciteSetBstMidEndSepPunct{\mcitedefaultmidpunct}
{\mcitedefaultendpunct}{\mcitedefaultseppunct}\relax
\EndOfBibitem
\bibitem[{\citenamefont{Chauve et~al.}(2000)\citenamefont{Chauve, Giamarchi,
  and Le~Doussal}}]{Chauve2000}
\bibinfo{author}{\bibfnamefont{P.}~\bibnamefont{Chauve}},
  \bibinfo{author}{\bibfnamefont{T.}~\bibnamefont{Giamarchi}},
  \bibnamefont{and}
  \bibinfo{author}{\bibfnamefont{P.}~\bibnamefont{Le~Doussal}},
  \bibinfo{journal}{Phys. Rev. B} \textbf{\bibinfo{volume}{62}},
  \bibinfo{pages}{6241} (\bibinfo{year}{2000})\relax
\mciteBstWouldAddEndPuncttrue
\mciteSetBstMidEndSepPunct{\mcitedefaultmidpunct}
{\mcitedefaultendpunct}{\mcitedefaultseppunct}\relax
\EndOfBibitem
\bibitem[{\citenamefont{Chauve et~al.}(2001)\citenamefont{Chauve, Le~Doussal,
  and J\"{o}rg~Wiese}}]{Chauve2001}
\bibinfo{author}{\bibfnamefont{P.}~\bibnamefont{Chauve}},
  \bibinfo{author}{\bibfnamefont{P.}~\bibnamefont{Le~Doussal}},
  \bibnamefont{and}
  \bibinfo{author}{\bibfnamefont{K.}~\bibnamefont{J\"{o}rg~Wiese}},
  \bibinfo{journal}{Phys. Rev. Lett.} \textbf{\bibinfo{volume}{86}},
  \bibinfo{pages}{1785} (\bibinfo{year}{2001})\relax
\mciteBstWouldAddEndPuncttrue
\mciteSetBstMidEndSepPunct{\mcitedefaultmidpunct}
{\mcitedefaultendpunct}{\mcitedefaultseppunct}\relax
\EndOfBibitem
\bibitem[{\citenamefont{Le~Doussal et~al.}(2002)\citenamefont{Le~Doussal,
  Wiese, and Chauve}}]{LeDoussal2002}
\bibinfo{author}{\bibfnamefont{P.}~\bibnamefont{Le~Doussal}},
  \bibinfo{author}{\bibfnamefont{K.~J.} \bibnamefont{Wiese}}, \bibnamefont{and}
  \bibinfo{author}{\bibfnamefont{P.}~\bibnamefont{Chauve}},
  \bibinfo{journal}{Phys. Rev. B} \textbf{\bibinfo{volume}{66}},
  \bibinfo{pages}{174201} (\bibinfo{year}{2002})\relax
\mciteBstWouldAddEndPuncttrue
\mciteSetBstMidEndSepPunct{\mcitedefaultmidpunct}
{\mcitedefaultendpunct}{\mcitedefaultseppunct}\relax
\EndOfBibitem
\bibitem[{\citenamefont{Duru\"{o}z et~al.}(1995)\citenamefont{Duru\"{o}z,
  Clarke, Marcus, and Harris}}]{Duruoz1995}
\bibinfo{author}{\bibfnamefont{C.~I.} \bibnamefont{Duru\"{o}z}},
  \bibinfo{author}{\bibfnamefont{R.~M.} \bibnamefont{Clarke}},
  \bibinfo{author}{\bibfnamefont{C.~M.} \bibnamefont{Marcus}},
  \bibnamefont{and} \bibinfo{author}{\bibfnamefont{J.~S.} \bibnamefont{Harris},
  \bibfnamefont{Jr.}}, \bibinfo{journal}{Phys. Rev. Lett.}
  \textbf{\bibinfo{volume}{74}}, \bibinfo{pages}{3237}
  (\bibinfo{year}{1995})\relax
\mciteBstWouldAddEndPuncttrue
\mciteSetBstMidEndSepPunct{\mcitedefaultmidpunct}
{\mcitedefaultendpunct}{\mcitedefaultseppunct}\relax
\EndOfBibitem
\bibitem[{\citenamefont{Rimberg et~al.}(1995)\citenamefont{Rimberg, Ho, and
  Clarke}}]{Rimberg1995}
\bibinfo{author}{\bibfnamefont{A.~J.} \bibnamefont{Rimberg}},
  \bibinfo{author}{\bibfnamefont{T.~R.} \bibnamefont{Ho}}, \bibnamefont{and}
  \bibinfo{author}{\bibfnamefont{J.}~\bibnamefont{Clarke}},
  \bibinfo{journal}{Phys. Rev. Lett.} \textbf{\bibinfo{volume}{74}},
  \bibinfo{pages}{4714} (\bibinfo{year}{1995})\relax
\mciteBstWouldAddEndPuncttrue
\mciteSetBstMidEndSepPunct{\mcitedefaultmidpunct}
{\mcitedefaultendpunct}{\mcitedefaultseppunct}\relax
\EndOfBibitem
\bibitem[{\citenamefont{Kurdak et~al.}(1998)\citenamefont{Kurdak, Rimberg, Ho,
  and Clarke}}]{Kurdak1998}
\bibinfo{author}{\bibfnamefont{C.}~\bibnamefont{Kurdak}},
  \bibinfo{author}{\bibfnamefont{A.~J.} \bibnamefont{Rimberg}},
  \bibinfo{author}{\bibfnamefont{T.~R.} \bibnamefont{Ho}}, \bibnamefont{and}
  \bibinfo{author}{\bibfnamefont{J.}~\bibnamefont{Clarke}},
  \bibinfo{journal}{Phys. Rev. B} \textbf{\bibinfo{volume}{57}},
  \bibinfo{pages}{R6842} (\bibinfo{year}{1998})\relax
\mciteBstWouldAddEndPuncttrue
\mciteSetBstMidEndSepPunct{\mcitedefaultmidpunct}
{\mcitedefaultendpunct}{\mcitedefaultseppunct}\relax
\EndOfBibitem
\bibitem[{\citenamefont{Parthasarathy et~al.}(2001)\citenamefont{Parthasarathy,
  Lin, and Jaeger}}]{Parthasarathy2001}
\bibinfo{author}{\bibfnamefont{R.}~\bibnamefont{Parthasarathy}},
  \bibinfo{author}{\bibfnamefont{X.-M.} \bibnamefont{Lin}}, \bibnamefont{and}
  \bibinfo{author}{\bibfnamefont{H.~M.} \bibnamefont{Jaeger}},
  \bibinfo{journal}{Phys. Rev. Lett.} \textbf{\bibinfo{volume}{87}},
  \bibinfo{pages}{186807} (\bibinfo{year}{2001})\relax
\mciteBstWouldAddEndPuncttrue
\mciteSetBstMidEndSepPunct{\mcitedefaultmidpunct}
{\mcitedefaultendpunct}{\mcitedefaultseppunct}\relax
\EndOfBibitem
\bibitem[{\citenamefont{Higgins and Bhattacharya}(1996)}]{Higgins1996}
\bibinfo{author}{\bibfnamefont{M.~J.} \bibnamefont{Higgins}} \bibnamefont{and}
  \bibinfo{author}{\bibfnamefont{S.}~\bibnamefont{Bhattacharya}},
  \bibinfo{journal}{Physica C: Superconductivity}
  \textbf{\bibinfo{volume}{257}}, \bibinfo{pages}{232} (\bibinfo{year}{1996}),
  ISSN \bibinfo{issn}{0921-4534}\relax
\mciteBstWouldAddEndPuncttrue
\mciteSetBstMidEndSepPunct{\mcitedefaultmidpunct}
{\mcitedefaultendpunct}{\mcitedefaultseppunct}\relax
\EndOfBibitem
\bibitem[{\citenamefont{Ruyter et~al.}(2008)\citenamefont{Ruyter, Plessis,
  Simon, Wahl, and Ammor}}]{Ruyter2008}
\bibinfo{author}{\bibfnamefont{A.}~\bibnamefont{Ruyter}},
  \bibinfo{author}{\bibfnamefont{D.}~\bibnamefont{Plessis}},
  \bibinfo{author}{\bibfnamefont{C.}~\bibnamefont{Simon}},
  \bibinfo{author}{\bibfnamefont{A.}~\bibnamefont{Wahl}}, \bibnamefont{and}
  \bibinfo{author}{\bibfnamefont{L.}~\bibnamefont{Ammor}},
  \bibinfo{journal}{Phys. Rev. B} \textbf{\bibinfo{volume}{77}},
  \bibinfo{pages}{212507} (\bibinfo{year}{2008})\relax
\mciteBstWouldAddEndPuncttrue
\mciteSetBstMidEndSepPunct{\mcitedefaultmidpunct}
{\mcitedefaultendpunct}{\mcitedefaultseppunct}\relax
\EndOfBibitem
\bibitem[{\citenamefont{Ammor et~al.}(2010)\citenamefont{Ammor, Ruyter,
  Shaidiuk, Hong, and Plessis}}]{Ammor2010}
\bibinfo{author}{\bibfnamefont{L.}~\bibnamefont{Ammor}},
  \bibinfo{author}{\bibfnamefont{A.}~\bibnamefont{Ruyter}},
  \bibinfo{author}{\bibfnamefont{V.~A.} \bibnamefont{Shaidiuk}},
  \bibinfo{author}{\bibfnamefont{N.~H.} \bibnamefont{Hong}}, \bibnamefont{and}
  \bibinfo{author}{\bibfnamefont{D.}~\bibnamefont{Plessis}},
  \bibinfo{journal}{Phys. Rev. B} \textbf{\bibinfo{volume}{81}},
  \bibinfo{pages}{094521} (\bibinfo{year}{2010})\relax
\mciteBstWouldAddEndPuncttrue
\mciteSetBstMidEndSepPunct{\mcitedefaultmidpunct}
{\mcitedefaultendpunct}{\mcitedefaultseppunct}\relax
\EndOfBibitem
\bibitem[{\citenamefont{Mohan et~al.}(2009)\citenamefont{Mohan, Sinha,
  Banerjee, Sood, Ramakrishnan, and Grover}}]{Mohan2009}
\bibinfo{author}{\bibfnamefont{S.}~\bibnamefont{Mohan}},
  \bibinfo{author}{\bibfnamefont{J.}~\bibnamefont{Sinha}},
  \bibinfo{author}{\bibfnamefont{S.~S.} \bibnamefont{Banerjee}},
  \bibinfo{author}{\bibfnamefont{A.~K.} \bibnamefont{Sood}},
  \bibinfo{author}{\bibfnamefont{S.}~\bibnamefont{Ramakrishnan}},
  \bibnamefont{and} \bibinfo{author}{\bibfnamefont{A.~K.}
  \bibnamefont{Grover}}, \bibinfo{journal}{Phys. Rev. Lett.}
  \textbf{\bibinfo{volume}{103}}, \bibinfo{pages}{167001}
  (\bibinfo{year}{2009})\relax
\mciteBstWouldAddEndPuncttrue
\mciteSetBstMidEndSepPunct{\mcitedefaultmidpunct}
{\mcitedefaultendpunct}{\mcitedefaultseppunct}\relax
\EndOfBibitem
\bibitem[{\citenamefont{Dom\'{\i}nguez}(1994)}]{Dominguez1994a}
\bibinfo{author}{\bibfnamefont{D.}~\bibnamefont{Dom\'{\i}nguez}},
  \bibinfo{journal}{Phys. Rev. Lett.} \textbf{\bibinfo{volume}{72}},
  \bibinfo{pages}{3096} (\bibinfo{year}{1994})\relax
\mciteBstWouldAddEndPuncttrue
\mciteSetBstMidEndSepPunct{\mcitedefaultmidpunct}
{\mcitedefaultendpunct}{\mcitedefaultseppunct}\relax
\EndOfBibitem
\bibitem[{\citenamefont{Chen et~al.}(2008)\citenamefont{Chen, Lv, and
  Liu}}]{Chen2008a}
\bibinfo{author}{\bibfnamefont{Q.-H.} \bibnamefont{Chen}},
  \bibinfo{author}{\bibfnamefont{J.-P.} \bibnamefont{Lv}}, \bibnamefont{and}
  \bibinfo{author}{\bibfnamefont{H.}~\bibnamefont{Liu}},
  \bibinfo{journal}{Phys. Rev. B} \textbf{\bibinfo{volume}{78}},
  \bibinfo{pages}{054519} (\bibinfo{year}{2008})\relax
\mciteBstWouldAddEndPuncttrue
\mciteSetBstMidEndSepPunct{\mcitedefaultmidpunct}
{\mcitedefaultendpunct}{\mcitedefaultseppunct}\relax
\EndOfBibitem
\bibitem[{\citenamefont{Chen}(2008)}]{Chen2008b}
\bibinfo{author}{\bibfnamefont{Q.-H.} \bibnamefont{Chen}},
  \bibinfo{journal}{Phys. Rev. B} \textbf{\bibinfo{volume}{78}},
  \bibinfo{pages}{104501} (\bibinfo{year}{2008})\relax
\mciteBstWouldAddEndPuncttrue
\mciteSetBstMidEndSepPunct{\mcitedefaultmidpunct}
{\mcitedefaultendpunct}{\mcitedefaultseppunct}\relax
\EndOfBibitem
\bibitem[{\citenamefont{Liu et~al.}(2008)\citenamefont{Liu, Zhou, and
  Chen}}]{Liu2008}
\bibinfo{author}{\bibfnamefont{H.}~\bibnamefont{Liu}},
  \bibinfo{author}{\bibfnamefont{W.}~\bibnamefont{Zhou}}, \bibnamefont{and}
  \bibinfo{author}{\bibfnamefont{Q.-H.} \bibnamefont{Chen}},
  \bibinfo{journal}{Phys. Rev. B} \textbf{\bibinfo{volume}{78}},
  \bibinfo{pages}{054509} (\bibinfo{year}{2008})\relax
\mciteBstWouldAddEndPuncttrue
\mciteSetBstMidEndSepPunct{\mcitedefaultmidpunct}
{\mcitedefaultendpunct}{\mcitedefaultseppunct}\relax
\EndOfBibitem
\bibitem[{\citenamefont{Guo et~al.}(2009)\citenamefont{Guo, Peng, and
  Chen}}]{Guo2009}
\bibinfo{author}{\bibfnamefont{F.}~\bibnamefont{Guo}, \bibfnamefont{Y.}},
  \bibinfo{author}{\bibfnamefont{L.}~\bibnamefont{Peng}, \bibfnamefont{H.}},
  \bibnamefont{and} \bibinfo{author}{\bibfnamefont{H.}~\bibnamefont{Chen},
  \bibfnamefont{Q.}}, \bibinfo{journal}{Eur. Phys. J. B}
  \textbf{\bibinfo{volume}{72}}, \bibinfo{pages}{591}
  (\bibinfo{year}{2009})\relax
\mciteBstWouldAddEndPuncttrue
\mciteSetBstMidEndSepPunct{\mcitedefaultmidpunct}
{\mcitedefaultendpunct}{\mcitedefaultseppunct}\relax
\EndOfBibitem
\bibitem[{\citenamefont{Lv et~al.}(2009)\citenamefont{Lv, Liu, and
  Chen}}]{Lv2009}
\bibinfo{author}{\bibfnamefont{J.-P.} \bibnamefont{Lv}},
  \bibinfo{author}{\bibfnamefont{H.}~\bibnamefont{Liu}}, \bibnamefont{and}
  \bibinfo{author}{\bibfnamefont{Q.-H.} \bibnamefont{Chen}},
  \bibinfo{journal}{Phys. Rev. B} \textbf{\bibinfo{volume}{79}},
  \bibinfo{pages}{104512} (\bibinfo{year}{2009})\relax
\mciteBstWouldAddEndPuncttrue
\mciteSetBstMidEndSepPunct{\mcitedefaultmidpunct}
{\mcitedefaultendpunct}{\mcitedefaultseppunct}\relax
\EndOfBibitem
\bibitem[{\citenamefont{Reichhardt et~al.}(2001)\citenamefont{Reichhardt,
  Olson, Gr\o{}nbech-Jensen, and Nori}}]{Reichhardt2001}
\bibinfo{author}{\bibfnamefont{C.}~\bibnamefont{Reichhardt}},
  \bibinfo{author}{\bibfnamefont{C.~J.} \bibnamefont{Olson}},
  \bibinfo{author}{\bibfnamefont{N.}~\bibnamefont{Gr\o{}nbech-Jensen}},
  \bibnamefont{and} \bibinfo{author}{\bibfnamefont{F.}~\bibnamefont{Nori}},
  \bibinfo{journal}{Phys. Rev. Lett.} \textbf{\bibinfo{volume}{86}},
  \bibinfo{pages}{4354} (\bibinfo{year}{2001})\relax
\mciteBstWouldAddEndPuncttrue
\mciteSetBstMidEndSepPunct{\mcitedefaultmidpunct}
{\mcitedefaultendpunct}{\mcitedefaultseppunct}\relax
\EndOfBibitem
\bibitem[{\citenamefont{Reichhardt and Olson}(2002)}]{Reichhardt2002}
\bibinfo{author}{\bibfnamefont{C.}~\bibnamefont{Reichhardt}} \bibnamefont{and}
  \bibinfo{author}{\bibfnamefont{C.~J.} \bibnamefont{Olson}},
  \bibinfo{journal}{Phys. Rev. Lett.} \textbf{\bibinfo{volume}{89}},
  \bibinfo{pages}{078301} (\bibinfo{year}{2002})\relax
\mciteBstWouldAddEndPuncttrue
\mciteSetBstMidEndSepPunct{\mcitedefaultmidpunct}
{\mcitedefaultendpunct}{\mcitedefaultseppunct}\relax
\EndOfBibitem
\bibitem[{\citenamefont{Reichhardt and
  Olson~Reichhardt}(2003)}]{Reichhardt2003}
\bibinfo{author}{\bibfnamefont{C.}~\bibnamefont{Reichhardt}} \bibnamefont{and}
  \bibinfo{author}{\bibfnamefont{C.~J.} \bibnamefont{Olson~Reichhardt}},
  \bibinfo{journal}{Phys. Rev. Lett.} \textbf{\bibinfo{volume}{90}},
  \bibinfo{pages}{046802} (\bibinfo{year}{2003})\relax
\mciteBstWouldAddEndPuncttrue
\mciteSetBstMidEndSepPunct{\mcitedefaultmidpunct}
{\mcitedefaultendpunct}{\mcitedefaultseppunct}\relax
\EndOfBibitem
\bibitem[{\citenamefont{Olive et~al.}(2009)\citenamefont{Olive, Fily, and
  Soret}}]{Olive2009}
\bibinfo{author}{\bibfnamefont{E.}~\bibnamefont{Olive}},
  \bibinfo{author}{\bibfnamefont{Y.}~\bibnamefont{Fily}}, \bibnamefont{and}
  \bibinfo{author}{\bibfnamefont{J.-C.} \bibnamefont{Soret}},
  \bibinfo{journal}{Journal of Physics: Conference Series}
  \textbf{\bibinfo{volume}{150}}, \bibinfo{pages}{052201}
  (\bibinfo{year}{2009}), ISSN \bibinfo{issn}{1742-6596}\relax
\mciteBstWouldAddEndPuncttrue
\mciteSetBstMidEndSepPunct{\mcitedefaultmidpunct}
{\mcitedefaultendpunct}{\mcitedefaultseppunct}\relax
\EndOfBibitem
\bibitem[{\citenamefont{Maeda et~al.}(1990)\citenamefont{Maeda, Notomi, and
  Uchinokura}}]{Maeda1990}
\bibinfo{author}{\bibfnamefont{A.}~\bibnamefont{Maeda}},
  \bibinfo{author}{\bibfnamefont{M.}~\bibnamefont{Notomi}}, \bibnamefont{and}
  \bibinfo{author}{\bibfnamefont{K.}~\bibnamefont{Uchinokura}},
  \bibinfo{journal}{Phys. Rev. B} \textbf{\bibinfo{volume}{42}},
  \bibinfo{pages}{3290} (\bibinfo{year}{1990})\relax
\mciteBstWouldAddEndPuncttrue
\mciteSetBstMidEndSepPunct{\mcitedefaultmidpunct}
{\mcitedefaultendpunct}{\mcitedefaultseppunct}\relax
\EndOfBibitem
\bibitem[{\citenamefont{Olson et~al.}(2001)\citenamefont{Olson, Reichhardt, and
  Vinokur}}]{Olson2001}
\bibinfo{author}{\bibfnamefont{C.~J.} \bibnamefont{Olson}},
  \bibinfo{author}{\bibfnamefont{C.}~\bibnamefont{Reichhardt}},
  \bibnamefont{and} \bibinfo{author}{\bibfnamefont{V.~M.}
  \bibnamefont{Vinokur}}, \bibinfo{journal}{Phys. Rev. B}
  \textbf{\bibinfo{volume}{64}}, \bibinfo{pages}{140502}
  (\bibinfo{year}{2001})\relax
\mciteBstWouldAddEndPuncttrue
\mciteSetBstMidEndSepPunct{\mcitedefaultmidpunct}
{\mcitedefaultendpunct}{\mcitedefaultseppunct}\relax
\EndOfBibitem
\bibitem[{\citenamefont{Marchetti et~al.}(2000)\citenamefont{Marchetti,
  Middleton, and Prellberg}}]{Marchetti2000}
\bibinfo{author}{\bibfnamefont{M.~C.} \bibnamefont{Marchetti}},
  \bibinfo{author}{\bibfnamefont{A.~A.} \bibnamefont{Middleton}},
  \bibnamefont{and}
  \bibinfo{author}{\bibfnamefont{T.}~\bibnamefont{Prellberg}},
  \bibinfo{journal}{Phys. Rev. Lett.} \textbf{\bibinfo{volume}{85}},
  \bibinfo{pages}{1104} (\bibinfo{year}{2000})\relax
\mciteBstWouldAddEndPuncttrue
\mciteSetBstMidEndSepPunct{\mcitedefaultmidpunct}
{\mcitedefaultendpunct}{\mcitedefaultseppunct}\relax
\EndOfBibitem
\bibitem[{\citenamefont{Marchetti and Saunders}(2002)}]{Marchetti2002}
\bibinfo{author}{\bibfnamefont{M.~C.} \bibnamefont{Marchetti}}
  \bibnamefont{and} \bibinfo{author}{\bibfnamefont{K.}~\bibnamefont{Saunders}},
  \bibinfo{journal}{Phys. Rev. B} \textbf{\bibinfo{volume}{66}},
  \bibinfo{pages}{224113} (\bibinfo{year}{2002})\relax
\mciteBstWouldAddEndPuncttrue
\mciteSetBstMidEndSepPunct{\mcitedefaultmidpunct}
{\mcitedefaultendpunct}{\mcitedefaultseppunct}\relax
\EndOfBibitem
\bibitem[{\citenamefont{Marchetti and Dahmen}(2002)}]{Marchetti2002a}
\bibinfo{author}{\bibfnamefont{M.~C.} \bibnamefont{Marchetti}}
  \bibnamefont{and} \bibinfo{author}{\bibfnamefont{K.~A.}
  \bibnamefont{Dahmen}}, \bibinfo{journal}{Phys. Rev. B}
  \textbf{\bibinfo{volume}{66}}, \bibinfo{pages}{214201}
  (\bibinfo{year}{2002})\relax
\mciteBstWouldAddEndPuncttrue
\mciteSetBstMidEndSepPunct{\mcitedefaultmidpunct}
{\mcitedefaultendpunct}{\mcitedefaultseppunct}\relax
\EndOfBibitem
\bibitem[{\citenamefont{Marchetti et~al.}(2003)\citenamefont{Marchetti,
  Middleton, Saunders, and Schwarz}}]{Marchetti2003}
\bibinfo{author}{\bibfnamefont{M.~C.} \bibnamefont{Marchetti}},
  \bibinfo{author}{\bibfnamefont{A.~A.} \bibnamefont{Middleton}},
  \bibinfo{author}{\bibfnamefont{K.}~\bibnamefont{Saunders}}, \bibnamefont{and}
  \bibinfo{author}{\bibfnamefont{J.~M.} \bibnamefont{Schwarz}},
  \bibinfo{journal}{Phys. Rev. Lett.} \textbf{\bibinfo{volume}{91}},
  \bibinfo{pages}{107002} (\bibinfo{year}{2003})\relax
\mciteBstWouldAddEndPuncttrue
\mciteSetBstMidEndSepPunct{\mcitedefaultmidpunct}
{\mcitedefaultendpunct}{\mcitedefaultseppunct}\relax
\EndOfBibitem
\bibitem[{\citenamefont{Marchetti}(2005)}]{Marchetti2005}
\bibinfo{author}{\bibfnamefont{M.}~\bibnamefont{Marchetti}},
  \bibinfo{journal}{Pramana} \textbf{\bibinfo{volume}{64}},
  \bibinfo{pages}{1097} (\bibinfo{year}{2005})\relax
\mciteBstWouldAddEndPuncttrue
\mciteSetBstMidEndSepPunct{\mcitedefaultmidpunct}
{\mcitedefaultendpunct}{\mcitedefaultseppunct}\relax
\EndOfBibitem
\bibitem[{\citenamefont{Le~Doussal et~al.}(2008)\citenamefont{Le~Doussal,
  Marchetti, and Wiese}}]{LeDoussal2008}
\bibinfo{author}{\bibfnamefont{P.}~\bibnamefont{Le~Doussal}},
  \bibinfo{author}{\bibfnamefont{M.~C.} \bibnamefont{Marchetti}},
  \bibnamefont{and} \bibinfo{author}{\bibfnamefont{K.~J.} \bibnamefont{Wiese}},
  \bibinfo{journal}{Phys. Rev. B} \textbf{\bibinfo{volume}{78}},
  \bibinfo{pages}{224201} (\bibinfo{year}{2008})\relax
\mciteBstWouldAddEndPuncttrue
\mciteSetBstMidEndSepPunct{\mcitedefaultmidpunct}
{\mcitedefaultendpunct}{\mcitedefaultseppunct}\relax
\EndOfBibitem
\bibitem[{\citenamefont{Strogatz et~al.}(1988)\citenamefont{Strogatz, Marcus,
  Westervelt, and Mirollo}}]{Strogatz1988}
\bibinfo{author}{\bibfnamefont{S.~H.} \bibnamefont{Strogatz}},
  \bibinfo{author}{\bibfnamefont{C.~M.} \bibnamefont{Marcus}},
  \bibinfo{author}{\bibfnamefont{R.~M.} \bibnamefont{Westervelt}},
  \bibnamefont{and} \bibinfo{author}{\bibfnamefont{R.~E.}
  \bibnamefont{Mirollo}}, \bibinfo{journal}{Phys. Rev. Lett.}
  \textbf{\bibinfo{volume}{61}}, \bibinfo{pages}{2380}
  (\bibinfo{year}{1988})\relax
\mciteBstWouldAddEndPuncttrue
\mciteSetBstMidEndSepPunct{\mcitedefaultmidpunct}
{\mcitedefaultendpunct}{\mcitedefaultseppunct}\relax
\EndOfBibitem
\bibitem[{\citenamefont{Levy et~al.}(1992)\citenamefont{Levy, Sherwin, Abraham,
  and Wiesenfeld}}]{Levy1992}
\bibinfo{author}{\bibfnamefont{J.}~\bibnamefont{Levy}},
  \bibinfo{author}{\bibfnamefont{M.~S.} \bibnamefont{Sherwin}},
  \bibinfo{author}{\bibfnamefont{F.~F.} \bibnamefont{Abraham}},
  \bibnamefont{and}
  \bibinfo{author}{\bibfnamefont{K.}~\bibnamefont{Wiesenfeld}},
  \bibinfo{journal}{Phys. Rev. Lett.} \textbf{\bibinfo{volume}{68}},
  \bibinfo{pages}{2968} (\bibinfo{year}{1992})\relax
\mciteBstWouldAddEndPuncttrue
\mciteSetBstMidEndSepPunct{\mcitedefaultmidpunct}
{\mcitedefaultendpunct}{\mcitedefaultseppunct}\relax
\EndOfBibitem
\bibitem[{\citenamefont{Watson and Fisher}(1996)}]{Watson1996}
\bibinfo{author}{\bibfnamefont{J.}~\bibnamefont{Watson}} \bibnamefont{and}
  \bibinfo{author}{\bibfnamefont{D.~S.} \bibnamefont{Fisher}},
  \bibinfo{journal}{Phys. Rev. B} \textbf{\bibinfo{volume}{54}},
  \bibinfo{pages}{938} (\bibinfo{year}{1996})\relax
\mciteBstWouldAddEndPuncttrue
\mciteSetBstMidEndSepPunct{\mcitedefaultmidpunct}
{\mcitedefaultendpunct}{\mcitedefaultseppunct}\relax
\EndOfBibitem
\bibitem[{\citenamefont{Watson and Fisher}(1997)}]{Watson1997}
\bibinfo{author}{\bibfnamefont{J.}~\bibnamefont{Watson}} \bibnamefont{and}
  \bibinfo{author}{\bibfnamefont{D.~S.} \bibnamefont{Fisher}},
  \bibinfo{journal}{Phys. Rev. B} \textbf{\bibinfo{volume}{55}},
  \bibinfo{pages}{14909} (\bibinfo{year}{1997})\relax
\mciteBstWouldAddEndPuncttrue
\mciteSetBstMidEndSepPunct{\mcitedefaultmidpunct}
{\mcitedefaultendpunct}{\mcitedefaultseppunct}\relax
\EndOfBibitem
\bibitem[{\citenamefont{Nogawa et~al.}(2003)\citenamefont{Nogawa, Matsukawa,
  and Yoshino}}]{Nogawa2003}
\bibinfo{author}{\bibfnamefont{T.}~\bibnamefont{Nogawa}},
  \bibinfo{author}{\bibfnamefont{H.}~\bibnamefont{Matsukawa}},
  \bibnamefont{and} \bibinfo{author}{\bibfnamefont{H.}~\bibnamefont{Yoshino}},
  \bibinfo{journal}{Physica B: Condensed Matter}
  \textbf{\bibinfo{volume}{329-333}}, \bibinfo{pages}{1448}
  (\bibinfo{year}{2003}), ISSN \bibinfo{issn}{0921-4526}\relax
\mciteBstWouldAddEndPuncttrue
\mciteSetBstMidEndSepPunct{\mcitedefaultmidpunct}
{\mcitedefaultendpunct}{\mcitedefaultseppunct}\relax
\EndOfBibitem
\bibitem[{\citenamefont{Kawaguchi}(1999)}]{Kawaguchi1999}
\bibinfo{author}{\bibfnamefont{T.}~\bibnamefont{Kawaguchi}},
  \bibinfo{journal}{Physics Letters A} \textbf{\bibinfo{volume}{251}},
  \bibinfo{pages}{73} (\bibinfo{year}{1999}), ISSN
  \bibinfo{issn}{0375-9601}\relax
\mciteBstWouldAddEndPuncttrue
\mciteSetBstMidEndSepPunct{\mcitedefaultmidpunct}
{\mcitedefaultendpunct}{\mcitedefaultseppunct}\relax
\EndOfBibitem
\bibitem[{\citenamefont{Schwarz and Fisher}(2001)}]{Schwarz2001}
\bibinfo{author}{\bibfnamefont{J.~M.} \bibnamefont{Schwarz}} \bibnamefont{and}
  \bibinfo{author}{\bibfnamefont{D.~S.} \bibnamefont{Fisher}},
  \bibinfo{journal}{Phys. Rev. Lett.} \textbf{\bibinfo{volume}{87}},
  \bibinfo{pages}{096107} (\bibinfo{year}{2001})\relax
\mciteBstWouldAddEndPuncttrue
\mciteSetBstMidEndSepPunct{\mcitedefaultmidpunct}
{\mcitedefaultendpunct}{\mcitedefaultseppunct}\relax
\EndOfBibitem
\bibitem[{\citenamefont{Schwarz and Fisher}(2003)}]{Schwarz2003}
\bibinfo{author}{\bibfnamefont{J.~M.} \bibnamefont{Schwarz}} \bibnamefont{and}
  \bibinfo{author}{\bibfnamefont{D.~S.} \bibnamefont{Fisher}},
  \bibinfo{journal}{Phys. Rev. E} \textbf{\bibinfo{volume}{67}},
  \bibinfo{pages}{021603} (\bibinfo{year}{2003})\relax
\mciteBstWouldAddEndPuncttrue
\mciteSetBstMidEndSepPunct{\mcitedefaultmidpunct}
{\mcitedefaultendpunct}{\mcitedefaultseppunct}\relax
\EndOfBibitem
\bibitem[{\citenamefont{Sengupta et~al.}(2010)\citenamefont{Sengupta, Sengupta,
  and Menon}}]{Sengupta2010}
\bibinfo{author}{\bibfnamefont{A.}~\bibnamefont{Sengupta}},
  \bibinfo{author}{\bibfnamefont{S.}~\bibnamefont{Sengupta}}, \bibnamefont{and}
  \bibinfo{author}{\bibfnamefont{G.~I.} \bibnamefont{Menon}},
  \bibinfo{journal}{Phys. Rev. B} \textbf{\bibinfo{volume}{81}},
  \bibinfo{pages}{144521} (\bibinfo{year}{2010})\relax
\mciteBstWouldAddEndPuncttrue
\mciteSetBstMidEndSepPunct{\mcitedefaultmidpunct}
{\mcitedefaultendpunct}{\mcitedefaultseppunct}\relax
\EndOfBibitem
\bibitem[{\citenamefont{Jensen et~al.}(1988{\natexlab{a}})\citenamefont{Jensen,
  Brass, and Berlinsky}}]{Jensen1988}
\bibinfo{author}{\bibfnamefont{H.~J.} \bibnamefont{Jensen}},
  \bibinfo{author}{\bibfnamefont{A.}~\bibnamefont{Brass}}, \bibnamefont{and}
  \bibinfo{author}{\bibfnamefont{A.~J.} \bibnamefont{Berlinsky}},
  \bibinfo{journal}{Phys. Rev. Lett.} \textbf{\bibinfo{volume}{60}},
  \bibinfo{pages}{1676} (\bibinfo{year}{1988}{\natexlab{a}})\relax
\mciteBstWouldAddEndPuncttrue
\mciteSetBstMidEndSepPunct{\mcitedefaultmidpunct}
{\mcitedefaultendpunct}{\mcitedefaultseppunct}\relax
\EndOfBibitem
\bibitem[{\citenamefont{Jensen et~al.}(1988{\natexlab{b}})\citenamefont{Jensen,
  Brass, Brechet, and Berlinsky}}]{Jensen1988a}
\bibinfo{author}{\bibfnamefont{H.~J.} \bibnamefont{Jensen}},
  \bibinfo{author}{\bibfnamefont{A.}~\bibnamefont{Brass}},
  \bibinfo{author}{\bibfnamefont{Y.}~\bibnamefont{Brechet}}, \bibnamefont{and}
  \bibinfo{author}{\bibfnamefont{A.~J.} \bibnamefont{Berlinsky}},
  \bibinfo{journal}{Phys. Rev. B} \textbf{\bibinfo{volume}{38}},
  \bibinfo{pages}{9235} (\bibinfo{year}{1988}{\natexlab{b}})\relax
\mciteBstWouldAddEndPuncttrue
\mciteSetBstMidEndSepPunct{\mcitedefaultmidpunct}
{\mcitedefaultendpunct}{\mcitedefaultseppunct}\relax
\EndOfBibitem
\bibitem[{\citenamefont{Koshelev}(1992)}]{Koshelev1992}
\bibinfo{author}{\bibfnamefont{A.}~\bibnamefont{Koshelev}},
  \bibinfo{journal}{Physica C: Superconductivity}
  \textbf{\bibinfo{volume}{198}}, \bibinfo{pages}{371} (\bibinfo{year}{1992}),
  ISSN \bibinfo{issn}{0921-4534}\relax
\mciteBstWouldAddEndPuncttrue
\mciteSetBstMidEndSepPunct{\mcitedefaultmidpunct}
{\mcitedefaultendpunct}{\mcitedefaultseppunct}\relax
\EndOfBibitem
\bibitem[{\citenamefont{Ryu et~al.}(1996)\citenamefont{Ryu, Hellerqvist,
  Doniach, Kapitulnik, and Stroud}}]{Ryu1996}
\bibinfo{author}{\bibfnamefont{S.}~\bibnamefont{Ryu}},
  \bibinfo{author}{\bibfnamefont{M.}~\bibnamefont{Hellerqvist}},
  \bibinfo{author}{\bibfnamefont{S.}~\bibnamefont{Doniach}},
  \bibinfo{author}{\bibfnamefont{A.}~\bibnamefont{Kapitulnik}},
  \bibnamefont{and} \bibinfo{author}{\bibfnamefont{D.}~\bibnamefont{Stroud}},
  \bibinfo{journal}{Phys. Rev. Lett.} \textbf{\bibinfo{volume}{77}},
  \bibinfo{pages}{5114} (\bibinfo{year}{1996})\relax
\mciteBstWouldAddEndPuncttrue
\mciteSetBstMidEndSepPunct{\mcitedefaultmidpunct}
{\mcitedefaultendpunct}{\mcitedefaultseppunct}\relax
\EndOfBibitem
\bibitem[{\citenamefont{Moon et~al.}(1996)\citenamefont{Moon, Scalettar, and
  Zim\'anyi}}]{Moon1996}
\bibinfo{author}{\bibfnamefont{K.}~\bibnamefont{Moon}},
  \bibinfo{author}{\bibfnamefont{R.~T.} \bibnamefont{Scalettar}},
  \bibnamefont{and} \bibinfo{author}{\bibfnamefont{G.~T.}
  \bibnamefont{Zim\'anyi}}, \bibinfo{journal}{Phys. Rev. Lett.}
  \textbf{\bibinfo{volume}{77}}, \bibinfo{pages}{2778}
  (\bibinfo{year}{1996})\relax
\mciteBstWouldAddEndPuncttrue
\mciteSetBstMidEndSepPunct{\mcitedefaultmidpunct}
{\mcitedefaultendpunct}{\mcitedefaultseppunct}\relax
\EndOfBibitem
\bibitem[{\citenamefont{Faleski et~al.}(1996)\citenamefont{Faleski, Marchetti,
  and Middleton}}]{Faleski1996}
\bibinfo{author}{\bibfnamefont{M.~C.} \bibnamefont{Faleski}},
  \bibinfo{author}{\bibfnamefont{M.~C.} \bibnamefont{Marchetti}},
  \bibnamefont{and} \bibinfo{author}{\bibfnamefont{A.~A.}
  \bibnamefont{Middleton}}, \bibinfo{journal}{Phys. Rev. B}
  \textbf{\bibinfo{volume}{54}}, \bibinfo{pages}{12427}
  (\bibinfo{year}{1996})\relax
\mciteBstWouldAddEndPuncttrue
\mciteSetBstMidEndSepPunct{\mcitedefaultmidpunct}
{\mcitedefaultendpunct}{\mcitedefaultseppunct}\relax
\EndOfBibitem
\bibitem[{\citenamefont{Spencer and Jensen}(1997)}]{Spencer1997}
\bibinfo{author}{\bibfnamefont{S.}~\bibnamefont{Spencer}} \bibnamefont{and}
  \bibinfo{author}{\bibfnamefont{H.~J.} \bibnamefont{Jensen}},
  \bibinfo{journal}{Phys. Rev. B} \textbf{\bibinfo{volume}{55}},
  \bibinfo{pages}{8473} (\bibinfo{year}{1997})\relax
\mciteBstWouldAddEndPuncttrue
\mciteSetBstMidEndSepPunct{\mcitedefaultmidpunct}
{\mcitedefaultendpunct}{\mcitedefaultseppunct}\relax
\EndOfBibitem
\bibitem[{\citenamefont{Olson et~al.}(1998)\citenamefont{Olson, Reichhardt, and
  Nori}}]{Olson1998}
\bibinfo{author}{\bibfnamefont{C.~J.} \bibnamefont{Olson}},
  \bibinfo{author}{\bibfnamefont{C.}~\bibnamefont{Reichhardt}},
  \bibnamefont{and} \bibinfo{author}{\bibfnamefont{F.}~\bibnamefont{Nori}},
  \bibinfo{journal}{Phys. Rev. Lett.} \textbf{\bibinfo{volume}{81}},
  \bibinfo{pages}{3757} (\bibinfo{year}{1998})\relax
\mciteBstWouldAddEndPuncttrue
\mciteSetBstMidEndSepPunct{\mcitedefaultmidpunct}
{\mcitedefaultendpunct}{\mcitedefaultseppunct}\relax
\EndOfBibitem
\bibitem[{\citenamefont{Kolton et~al.}(1999)\citenamefont{Kolton,
  Dom\'{\i}nguez, and Gr\o{}nbech-Jensen}}]{Kolton1999}
\bibinfo{author}{\bibfnamefont{A.~B.} \bibnamefont{Kolton}},
  \bibinfo{author}{\bibfnamefont{D.}~\bibnamefont{Dom\'{\i}nguez}},
  \bibnamefont{and}
  \bibinfo{author}{\bibfnamefont{N.}~\bibnamefont{Gr\o{}nbech-Jensen}},
  \bibinfo{journal}{Phys. Rev. Lett.} \textbf{\bibinfo{volume}{83}},
  \bibinfo{pages}{3061} (\bibinfo{year}{1999})\relax
\mciteBstWouldAddEndPuncttrue
\mciteSetBstMidEndSepPunct{\mcitedefaultmidpunct}
{\mcitedefaultendpunct}{\mcitedefaultseppunct}\relax
\EndOfBibitem
\bibitem[{\citenamefont{Cao et~al.}(2000)\citenamefont{Cao, Jiao, and
  Ying}}]{Cao2000}
\bibinfo{author}{\bibfnamefont{Y.}~\bibnamefont{Cao}},
  \bibinfo{author}{\bibfnamefont{Z.}~\bibnamefont{Jiao}}, \bibnamefont{and}
  \bibinfo{author}{\bibfnamefont{H.}~\bibnamefont{Ying}},
  \bibinfo{journal}{Phys. Rev. B} \textbf{\bibinfo{volume}{62}},
  \bibinfo{pages}{4163} (\bibinfo{year}{2000})\relax
\mciteBstWouldAddEndPuncttrue
\mciteSetBstMidEndSepPunct{\mcitedefaultmidpunct}
{\mcitedefaultendpunct}{\mcitedefaultseppunct}\relax
\EndOfBibitem
\bibitem[{\citenamefont{Fangohr et~al.}(2001)\citenamefont{Fangohr, Cox, and
  de~Groot}}]{Fangohr2001}
\bibinfo{author}{\bibfnamefont{H.}~\bibnamefont{Fangohr}},
  \bibinfo{author}{\bibfnamefont{S.~J.} \bibnamefont{Cox}}, \bibnamefont{and}
  \bibinfo{author}{\bibfnamefont{P.~A.~J.} \bibnamefont{de~Groot}},
  \bibinfo{journal}{Phys. Rev. B} \textbf{\bibinfo{volume}{64}},
  \bibinfo{pages}{064505} (\bibinfo{year}{2001})\relax
\mciteBstWouldAddEndPuncttrue
\mciteSetBstMidEndSepPunct{\mcitedefaultmidpunct}
{\mcitedefaultendpunct}{\mcitedefaultseppunct}\relax
\EndOfBibitem
\bibitem[{\citenamefont{Chen et~al.}(2003)\citenamefont{Chen, Cao, and
  Jiao}}]{Chen2003}
\bibinfo{author}{\bibfnamefont{J.}~\bibnamefont{Chen}},
  \bibinfo{author}{\bibfnamefont{Y.}~\bibnamefont{Cao}}, \bibnamefont{and}
  \bibinfo{author}{\bibfnamefont{Z.}~\bibnamefont{Jiao}},
  \bibinfo{journal}{Physics Letters A} \textbf{\bibinfo{volume}{318}},
  \bibinfo{pages}{146} (\bibinfo{year}{2003}), ISSN
  \bibinfo{issn}{0375-9601}\relax
\mciteBstWouldAddEndPuncttrue
\mciteSetBstMidEndSepPunct{\mcitedefaultmidpunct}
{\mcitedefaultendpunct}{\mcitedefaultseppunct}\relax
\EndOfBibitem
\bibitem[{\citenamefont{Chandran et~al.}(2003)\citenamefont{Chandran,
  Scalettar, and Zim\'anyi}}]{Chandran2003}
\bibinfo{author}{\bibfnamefont{M.}~\bibnamefont{Chandran}},
  \bibinfo{author}{\bibfnamefont{R.~T.} \bibnamefont{Scalettar}},
  \bibnamefont{and} \bibinfo{author}{\bibfnamefont{G.~T.}
  \bibnamefont{Zim\'anyi}}, \bibinfo{journal}{Phys. Rev. B}
  \textbf{\bibinfo{volume}{67}}, \bibinfo{pages}{052507}
  (\bibinfo{year}{2003})\relax
\mciteBstWouldAddEndPuncttrue
\mciteSetBstMidEndSepPunct{\mcitedefaultmidpunct}
{\mcitedefaultendpunct}{\mcitedefaultseppunct}\relax
\EndOfBibitem
\bibitem[{\citenamefont{Gr\o{}nbech-Jensen
  et~al.}(1996)\citenamefont{Gr\o{}nbech-Jensen, Bishop, and
  Dom\'{\i}nguez}}]{Groenbech-Jensen1996}
\bibinfo{author}{\bibfnamefont{N.}~\bibnamefont{Gr\o{}nbech-Jensen}},
  \bibinfo{author}{\bibfnamefont{A.~R.} \bibnamefont{Bishop}},
  \bibnamefont{and}
  \bibinfo{author}{\bibfnamefont{D.}~\bibnamefont{Dom\'{\i}nguez}},
  \bibinfo{journal}{Phys. Rev. Lett.} \textbf{\bibinfo{volume}{76}},
  \bibinfo{pages}{2985} (\bibinfo{year}{1996})\relax
\mciteBstWouldAddEndPuncttrue
\mciteSetBstMidEndSepPunct{\mcitedefaultmidpunct}
{\mcitedefaultendpunct}{\mcitedefaultseppunct}\relax
\EndOfBibitem
\bibitem[{\citenamefont{Olive and Soret}(2006)}]{Olive2006}
\bibinfo{author}{\bibfnamefont{E.}~\bibnamefont{Olive}} \bibnamefont{and}
  \bibinfo{author}{\bibfnamefont{J.~C.} \bibnamefont{Soret}},
  \bibinfo{journal}{Phys. Rev. Lett.} \textbf{\bibinfo{volume}{96}},
  \bibinfo{pages}{027002} (\bibinfo{year}{2006})\relax
\mciteBstWouldAddEndPuncttrue
\mciteSetBstMidEndSepPunct{\mcitedefaultmidpunct}
{\mcitedefaultendpunct}{\mcitedefaultseppunct}\relax
\EndOfBibitem
\bibitem[{\citenamefont{Olive and Soret}(2008)}]{Olive2008}
\bibinfo{author}{\bibfnamefont{E.}~\bibnamefont{Olive}} \bibnamefont{and}
  \bibinfo{author}{\bibfnamefont{J.~C.} \bibnamefont{Soret}},
  \bibinfo{journal}{Phys. Rev. B} \textbf{\bibinfo{volume}{77}},
  \bibinfo{pages}{144514} (\bibinfo{year}{2008})\relax
\mciteBstWouldAddEndPuncttrue
\mciteSetBstMidEndSepPunct{\mcitedefaultmidpunct}
{\mcitedefaultendpunct}{\mcitedefaultseppunct}\relax
\EndOfBibitem
\bibitem[{\citenamefont{Reichhardt and Reichhardt}(2009)}]{Reichhardt2009}
\bibinfo{author}{\bibfnamefont{C.}~\bibnamefont{Reichhardt}} \bibnamefont{and}
  \bibinfo{author}{\bibfnamefont{C.~J.~O.} \bibnamefont{Reichhardt}},
  \bibinfo{journal}{Phys. Rev. Lett.} \textbf{\bibinfo{volume}{103}},
  \bibinfo{pages}{168301} (\bibinfo{year}{2009})\relax
\mciteBstWouldAddEndPuncttrue
\mciteSetBstMidEndSepPunct{\mcitedefaultmidpunct}
{\mcitedefaultendpunct}{\mcitedefaultseppunct}\relax
\EndOfBibitem
\bibitem[{\citenamefont{Luo and Hu}(2007)}]{Luo2007}
\bibinfo{author}{\bibfnamefont{M.-B.} \bibnamefont{Luo}} \bibnamefont{and}
  \bibinfo{author}{\bibfnamefont{X.}~\bibnamefont{Hu}}, \bibinfo{journal}{Phys.
  Rev. Lett.} \textbf{\bibinfo{volume}{98}}, \bibinfo{eid}{267002}
  (\bibinfo{year}{2007})\relax
\mciteBstWouldAddEndPuncttrue
\mciteSetBstMidEndSepPunct{\mcitedefaultmidpunct}
{\mcitedefaultendpunct}{\mcitedefaultseppunct}\relax
\EndOfBibitem
\bibitem[{\citenamefont{Middleton and Wingreen}(1993)}]{Middleton1993b}
\bibinfo{author}{\bibfnamefont{A.~A.} \bibnamefont{Middleton}}
  \bibnamefont{and} \bibinfo{author}{\bibfnamefont{N.~S.}
  \bibnamefont{Wingreen}}, \bibinfo{journal}{Phys. Rev. Lett.}
  \textbf{\bibinfo{volume}{71}}, \bibinfo{pages}{3198}
  (\bibinfo{year}{1993})\relax
\mciteBstWouldAddEndPuncttrue
\mciteSetBstMidEndSepPunct{\mcitedefaultmidpunct}
{\mcitedefaultendpunct}{\mcitedefaultseppunct}\relax
\EndOfBibitem
\end{mcitethebibliography}

\end{document}